\documentclass{layout/tudelft-aiaa}

\usepackage[T1]{fontenc}
\usepackage[utf8]{inputenc}

\usepackage{graphicx} 
\usepackage{amsmath} 
\usepackage{siunitx} 
\usepackage{tabularx} 
\usepackage{subcaption} 
\usepackage{rotating}
\usepackage{makecell}
\usepackage{adjustbox}
\usepackage{float}
\usepackage{lscape}
\usepackage{placeins}
\usepackage{comment}
\usepackage{setspace}
\title{Estimating the technical wind energy potential of Kansas that incorporates the atmospheric response for policy applications}

\usepackage[symbol]{footmisc}

\providecommand{\keywords}[1]
{
  \small	
  \textbf{\textit{Keywords---}} #1
}

\author[1]{Jonathan Minz\footnote{Corresponding author email: jminz@bgc-jena.mpg.de}}
\author[1]{Axel Kleidon}
\author[2,3]{Nsilulu T. Mbungu}
\author[4]{Lee M. Miller}
\affil[1]{Biospheric Theory and Modelling Group, Max Planck Institute of Biogeochemistry,Jena, Germany}
\affil[2]{Department of Electrical Engineering, Tshwane University of Technology, Pretoria, South Africa}
\affil[3]{Department of Electrical Engineering, University of Sharjah, Sharjah, United Arab Emirates}
\affil[4]{Pacific Northwest National Laboratory, Richland, Washington, United States of America}

\begin{document}
\footnote{Corresponding author email: jminz@bgc-jena.mpg.de (JM)}
\AlwaysPagewidth{

\maketitle

\begin{abstract}
    \noindent Energy scenarios and transition  pathways need estimates of technical wind energy potentials. However, the standard policy-side approach uses observed wind speeds, thereby neglecting the effects of kinetic energy (KE) removal by the wind turbines that depletes the regional wind resource, lowers wind speeds, and reduces capacity factors. The standard approach therefore significantly overestimates the wind resource potential relative to estimates using numerical models of the atmosphere with interactive wind farm parameterizations.  Here, we test the extent to which these effects of KE removal can be accounted for by our KE Budget of the Atmosphere (KEBA) approach over Kansas in the central US, a region with a high wind energy resource.  We find that KEBA reproduces the simulated estimates within 10 - 11\%, which are 30 - 50\% lower than estimates using the standard approach.  We also evaluate important differences in the depletion of the wind resource between daytime and nighttime conditions, which are due to effects of stability. Our results indicate that the KEBA approach is a simple yet adequate approach to evaluating regional-scale wind resource potentials, and that resource depletion effects need to be accounted for at such scales in policy applications.
  
\end{abstract}
}

\keywords{Wind resource estimation, technical wind energy potentials, regional limits, Kinetic Energy Budget of the Atmosphere, wind speeds, boundary layer, stability}






\section{Introduction}

Estimates of technical wind energy potential are important for the design of energy transition pathways towards a future sustainable energy system \citep{IRENA_2019,WVBros_2019,GEA_2012,iea_2021}. These are theoretical estimates of energy generation from hypothetical wind turbine deployments over large regions available for wind energy development. The standard approach to estimating them is to force a single wind turbine's power curve with observed or modelled wind speeds at the hub-height of the turbine and then multiply this by the total number of wind turbines within a deployment area \citep{Hoogwijk_2004,Archer_2005,Lu_2009,SchallenbergRodriguez_2013,Eurek_2017,Enevoldsen_2019}. Higher wind speeds, higher turbine densities, and larger turbine rated capacities are expected to lead to a higher technical potential\citep{Wiser2016}. Attaining these technical wind resource potentials is underpinned by the implicit assumption that the removal of kinetic energy (KE) by the wind turbines does not affect wind speeds over the deployment region. Single factor adjustments of 10 - 20\% are used to correct for inter-turbine interactions within individual wind farms. This potential divided by the installed capacity of the deployment is used to calculate a metric of efficiency known as the capacity factor. This is used in the estimation of the economic cost of large scale wind turbine deployments\citep{Blanco2009,RAGHEB2017537}. Technical potentials and capacity factors are the most significant control on the cost of wind energy \citep{Blanco2009,RAGHEB2017537}. Thus, accurate estimates of technical potential and capacity factors are critical for the design of robust energy transition pathways. 
\par 

However, simulations with numerical models of the atmosphere show that the standard approach significantly overestimates technical potential and capacity factors when wind energy is intensively used at large scales \citep{Adams_2013,Miller_2015,Miller_2016, Volker_2017,Kleidon_2020,Jacobson2012}. This overestimation results from the standard approach implicitly neglecting the wind speed reduction by the regional-scale extraction of KE by the wind turbines.  To understand this reduction effect on regional scales, we need to look at how kinetic energy is generated and transported towards the surface before it can be extracted by wind turbines.

The winds of the large-scale circulation and their associated kinetic energy are predominantly generated in the free atmosphere by differences in potential energy due to differential solar radiative heating \citep{Piexoto1992, Kleidon_2021}.  This KE is transported vertically downwards into the boundary layer, the lowest layer of the atmosphere where most of the KE is dissipated \citep{stull_2009}.  The turbines extract some of the KE which would otherwise have been dissipated by surface friction.  Since the rate at which KE is transported into the boundary layer is limited, it leads to a fixed KE budget being available for driving movement within the boundary layer\citep{Kleidon_2020}. This means that the extraction of KE by a large number of wind turbines leads to less KE being available for the motion of the winds. As a result, larger rates of KE extraction from a fixed KE budget causes slower winds and reduced capacity factors \citep{Miller_2011}. This is supported by numerical weather model simulations, which show that mean electricity generation potentials from onshore deployments greater than 100 km$^2$ are limited to yields of about 1.1 W m$^{-2}$ of surface area \citep{Adams_2013,Miller_2016,Jacobson2012,Marvel2012,Gustavson1979, wang2010potential,wang2011potential,Volker_2017}, which contrast with standard estimates ranging from 2-6 W m$^{-2}$\citep{JACOBSON2011,Jacobson2012,Lu_2009,Archer_2005,edenhofer2011ipcc,capps2010estimated}. At the maximum generation potential, wind speeds are estimated to slow by 42\%, while capacity factors reduce by $\sim$ 50\% relative to the standard estimate \citep{Miller_2015,Volker_2017}. A mean of 1.1 W m$^{-2}$ implies electricity generation of $\sim$ 900 - 1900 TWh yr$^{-1}$ if all the available area for wind energy in a windy area like Kansas (100,000 - 200,000 km$^{-2}$) is covered with wind turbines. These generation potentials are about a third lower than the standard expectation of 2000 - 3000 TWh yr$^{-1}$ \citep{brown_2016,Lopez_2012}. Thus, the policy-side approach to technical potential estimation needs to incorporate the effects of wind speed reductions arising from limitations imposed by the atmospheric KE budgets.

A simple yet physical approach to deriving technical potential estimates that includes the effects of KE removal on wind speeds is to constrain the wind speeds and turbine yields with an explicitly defined KE budget of the atmospheric boundary layer. In this approach, known as the Kinetic Energy Budget of the Atmosphere (KEBA, \citet{Kleidon_2020}), first the budget available to the deployment is estimated from the sum of the vertical and horizontal KE fluxes over the deployment. The vertical component represents the KE input into the boundary layer from the free atmosphere while the horizontal component represents the boundary layer wind flow.  Both rates can be estimated from wind speed observations, but also depend on boundary layer height and surface friction. The reduction in wind speeds is estimated by accounting for the removal of KE from the budget. The slower wind speeds are then used to estimate turbine yields\citep{Kleidon_2020}. This approach has previously been shown to compare well against numerical weather forecasting simulations of wind turbine deployments in idealized onshore weather conditions\citep{Kleidon_2020} and in real weather conditions in offshore areas in the German Bight of the North Sea \citep{AGORA_2020}. The goal here is to test this approach further in a realistic onshore region to understand the limits of its application for it to be used in policy-side resource estimation. 


\begin{figure}[p]
    \centering
    \includegraphics[width=8.5cm]{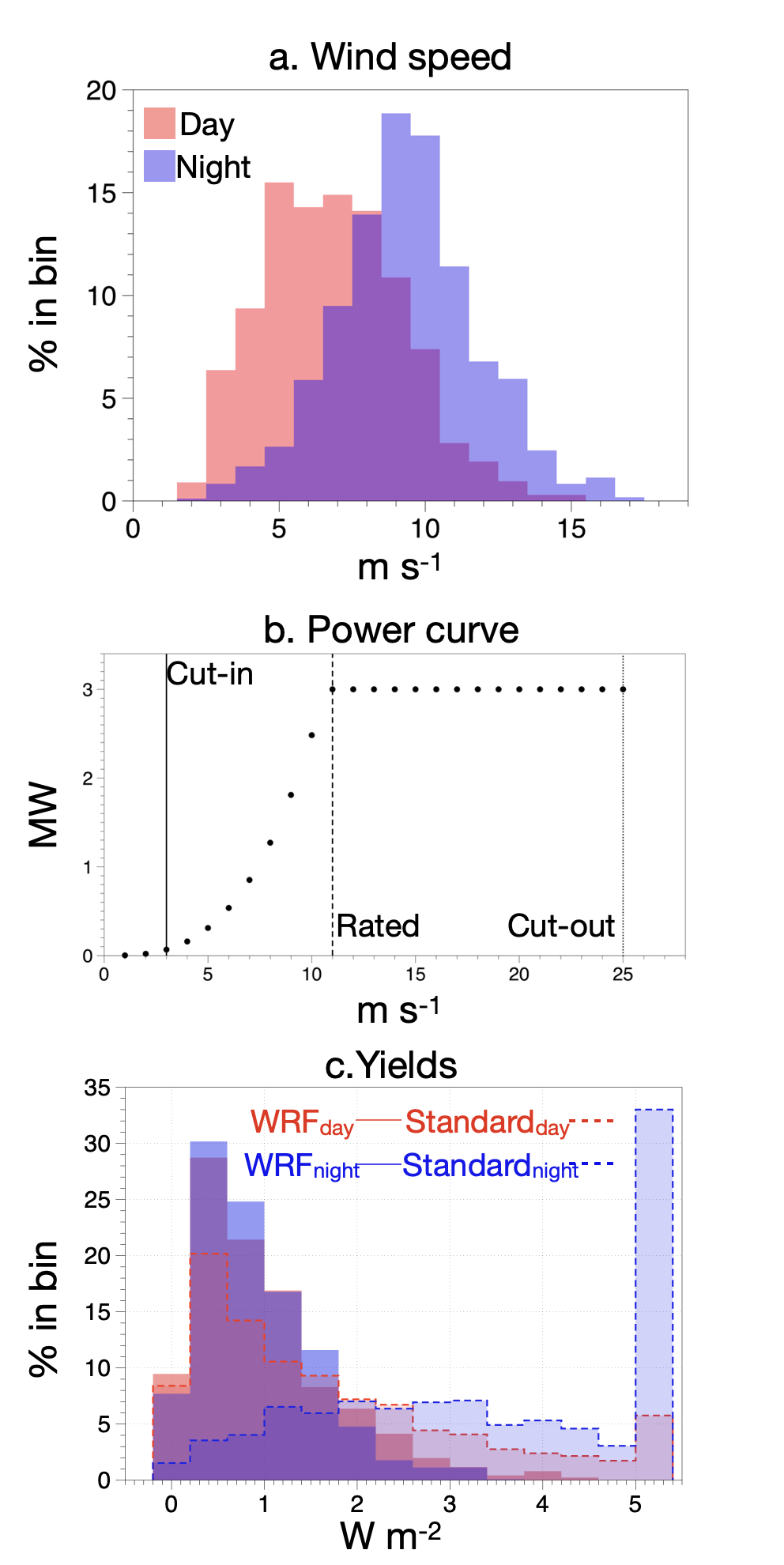}
    \caption{(a) Distribution of wind speeds averaged over a prospective deployment area in Kansas, central US, for daytime (red) and nighttime (blue) in the absence of wind turbines. (b) The power curve for a Vestas V112 3MW wind turbine used in this study. (c) Deployment yields during daytime (red) and nighttime (blue) for the 5 MW km$^{-2}$ scenario from an interactive WRF simulation (solid, "WRF"), and using the standard approach (dashed outlines). Data taken from \cite{Miller_2015}.}
    \label{fig:fig0}
\end{figure}

\begin{figure}[p!]
    \centering
    \includegraphics[width=16.5cm]{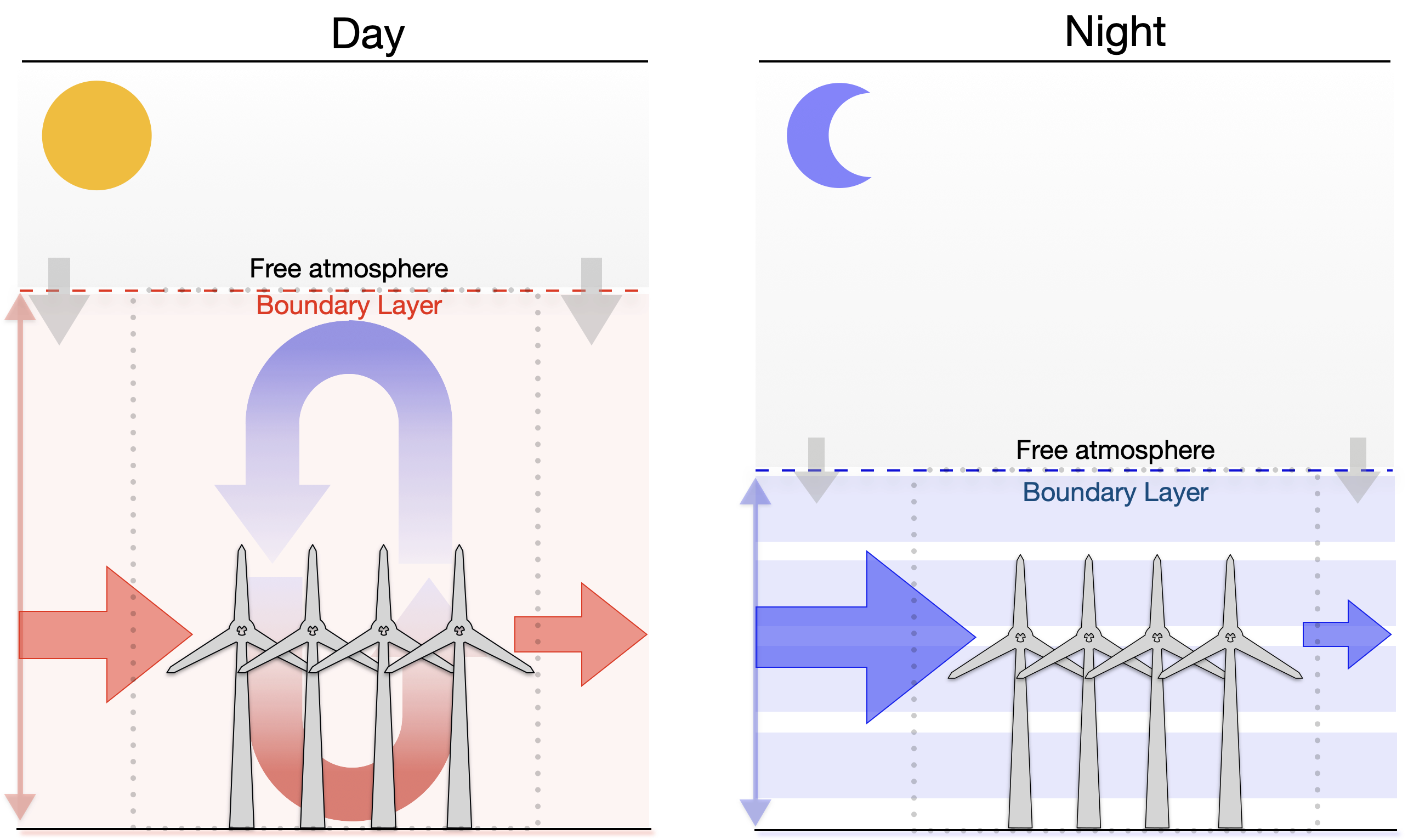}
    \caption{Differences between daytime (left) and nighttime (right) boundary layer conditions. Gray downward arrows represent the downward flux of kinetic energy from the free atmosphere into the boundary layer. The red and blue arrows represent the horizontal kinetic energy in- and outflow (from left to right) through the boundary layer volume bounding the regional scale wind turbine deployment (dotted box). The free atmosphere represents the part of the atmosphere in which large-scale motion is generated in the absence of friction.}
    \label{fig:fig1}
\end{figure}

In this study, we will use this approach to evaluate a seemingly counter-intuitive result (Fig. \ref{fig:fig0}) reported by previous numerical simulations of realistic large-scale wind turbine deployment scenarios in Kansas, central US, under realistic weather conditions \citep{Miller_2015}.  These simulations showed that wind speeds are typically 40\% lower during the day than at night (Fig \ref{fig:fig0}a), but overall daytime yields were about 50\% higher than nighttime (Fig \ref{fig:fig0}c). This is an important result to evaluate with KEBA since the standard approach would estimate the opposite, higher nighttime yields due to higher wind speeds. The result can be understood when one accounts for the effect of lower boundary layer heights and reduced mixing at night, which reduces the size of the kinetic energy budget (Fig. \ref{fig:fig1}).  As a result the KE removal by the wind turbines has a stronger effect on wind speed reductions at night \citep{Fitch2013b,Abkar2015}. During the day, because the KE budget is larger, this depletion effect is proportionally smaller. Solar insolation drives vertical convection and the vertical growth of the boundary layer, resulting in higher downward replenishment of KE from the free atmosphere and a larger reservoir of KE in the boundary layer. The absence of solar-driven convection at night leads to stratified or stable conditions that restricts vertical KE replenishment. This leads to greater reduction in wind speeds at night compared to the the day, and therefore lower yields, despite higher incoming, undisturbed wind speeds. Thus, the differences in boundary layer characteristics during day and night will affect wind resource potentials of regional deployments of wind turbines. 


Our goal in this paper is twofold: (1) to evaluate the effect just described to demonstrate the importance of a broader accounting of the atmosphere in resource potential estimation, and (2) to quantify the role of KE removal in shaping wind resource potentials at the regional scale.  To accomplish our first goal, we will explicitly evaluate this difference between day and night by evaluating separate daytime and nighttime budgets in the KEBA approach. This is accomplished by prescribing different boundary layer heights in KEBA, which changes the size of the KE resource.  Note that the effect of reduced turbulent mixing within the boundary layer is not accounted for.  To accomplish our second goal, we use the scenarios of the previously published study for Kansas \citep{Miller_2015}.  We use the wind speeds of their control simulation without deployment to evaluate wind resource potentials using the standard approach as well as KEBA, and compare these to the estimates derived from the interactive model simulations.  

In the following section, we provide a brief description of the KE budget approach, the turbine deployment scenarios, and the model parameters used. We then present the KE budgets diagnosed from the simulations for the different scenarios, the reductions in wind speeds with greater wind energy use, and describe the effects on yield estimates and capacity factors.  We also evaluate the importance of accounting for different boundary layer heights for the estimates.  After a brief account of limitations, we then re-evaluate the wind resource potential and compare it to previously-published estimates \citep{Lopez_2012,brown_2016}.  We close with a discussion and conclusions. 

\section{Methods}

We use the wind speeds, scenarions, and yield estimates from \citeauthor{Miller_2015}'s WRF simulations \citep{Miller_2015}. We use these simulations as the reference in which the effects of wind turbines on the atmosphere are fully accounted for, and refer to the yield estimate as the "WRF" estimate.  The simulations were performed with the WRF-ARW v3.3.1 regional weather forecasting model \citep{WRF_3_2009} to simulate different levels of hypothetical deployments of wind turbines over 112 $\cdot$ 10$^{3}$ km$^{2}$ in Kansas (Central US) using atmospheric conditions from May 15 to September 30, 2001. This period is considered to be climatologically representative for this region \citep{trier2010environmental}.  The deployment area is similar to that assumed in previous resource evaluations \citep{Lopez_2012,brown_2016}.  Wind turbines are parameterised as elevated momentum sinks and sources of additional Turbulent Kinetic Energy (TKE) \citep{Fitch_2013}.  The large, idealized deployments simulated a range of installed turbine capacity densities from 0.3125 to 100 MW km$^{-2}$ which were equally distributed within the expansive wind farm area.  Here we  restrict the comparison to a maximum installed capacity density of 10 MW km$^{-2}$, yielding a total installed capacity of 35 GW to 1.1 TW over the region. The turbine characteristics and wind park scenarios, as well as the symbols used in the following, are summarized in Tables \ref{table_01} and \ref{table_02}.  


\begin{table}
\centering
\caption{\label{label}Turbine characteristics of a Vestas V112 3 MW turbine, as in \cite{Miller_2015}.}
\begin{tabular}{@{}lllll}
\hline
\textbf{Description} &\textbf{Symbol} &\textbf{Value} &\textbf{Units} \\
\hline
 Hub-height & $H_{hub}$ & 84 & m \\
 Rotor diameter & $D$ & 112 & m \\
 Rotor area & $A_{rotor}$ & 9852 & m$^{2}$ \\
 Rated power  & $P_{el,max}$ & 3.075 & MW \\
 Cut-in wind speed  & $v_{min}$ & 3 & m s$^{-1}$ \\
 Rated wind speed  & $v_{rated}$ & 11.5 & m $^{-1}$ \\
 Cut-out wind speed  & $v_{max}$ & 25 & m s$^{-1}$ \\
 Power coefficient (max.) & $\eta_{max}$ & 0.42 & - \\
\hline
\end{tabular}
\label{table_01}
\end{table}


\begin{table}
\centering
\caption{\label{label}Scenarios of large-scale deployment of wind turbines in Kansas, Central US, evaluated here.  Based on \citep{Miller_2015}.}
\begin{tabular}{@{}lllll}
\hline
\textbf{Description} &\textbf{Symbol} &\textbf{Value} &\textbf{Units} \\
\hline
 Width & $W$ & $360 \cdot 10^{3}$ & m \\  
 Length & $L$ & $312 \cdot 10^{3}$ & m \\
 Capacity density & - & $0.3125 - 10$ & MW km$^{-2}$ \\
 Number of turbines & $N$ & $11.7 \times 10^{3}$ - $3.7\times 10^{5}$  & - \\
 Deployment area & $A_{farm}$ & $1.12 \times 10^{11} $ & m$^{2}$ \\
\hline
\end{tabular}
\label{table_02}
\end{table}

The "standard" and "KEBA" yield, or electricity, estimates were then calculated using hourly time series of wind speeds, $v_{in}$, from the WRF Control simulation, i.e., without any wind turbines present. The "standard" estimate replicates the standard approach used in existing policy side evaluations. It is based on the power curve of the turbine, the number of turbines in the scenario, and the wind speeds, $v_{in}$. The electricity yield of the standard approach, $P_{el,std}$, is estimated from the turbine's power curve (Fig. \ref{fig:fig0}b) by

\begin{equation}
\label{P_Common}
P_{el,std} = N \cdot \min (P_{el,max}, \dfrac{\rho}{2} \cdot \eta_{max} \cdot A_{rotor} \cdot v_{in}^{3})    
\end{equation}

where $N$ is the number of turbines, $P_{el,max}$ is the rated capacity of the turbine, $\rho$ is the air density (we used $\rho = 1.2$ kg m$^{-3}$), $\eta_{max}$ is the maximum power coefficient, and $A_{rotor}$ is the rotor-swept area, and $v_{in}$ is the wind speed from the WRF Control simulation (Table \ref{table_01}).  This estimate assumes that the effects of the KE removal by the wind turbines is fully compensated for by the inter-turbine spacing, which allows wind speeds and capacity factors to be unaffected by presence of a wind turbine deployment in the region.

The "KEBA" estimate is derived from the KEBA model \citep{Kleidon_2020} augmented with information about day- and nighttime boundary layer heights derived from the WRF simulations.  The budgeting of the KE fluxes of the boundary layer over the deployment region results in a reduction factor $f_{red}$. This factor encapsulates the effect of KE removal from the wind by the turbines on wind speeds and turbine yields. In KEBA, the wake loss term is fixed as half of the deployment yield after the work of \citeauthor{Corten_2001}\citep{Corten_2001}. First, $f_{red}$ is applied to the WRF control wind speeds ($v_{in}$) to quantify the reduction in the wind speeds ($v_{eff}$).

\begin{equation}
\label{keba_veff}
v_{eff} = f_{red}^{\frac{1}{3}} \cdot v_{in}   
\end{equation}

Then, $v_{eff}$ is applied to the standard estimate (Equation \ref{P_Common}) instead of $v_{in}$ to derive the KEBA estimate of deployment yields ($P_{el,keba}$). This results in the following expression for deployment yield (see \cite{Kleidon_2020} for detailed derivation)

\begin{equation}
    P_{el,keba} = N \cdot \min (P_{el,max}, f_{red} \cdot \dfrac{\rho}{2} \cdot \eta_{max} \cdot A_{rotor} \cdot v_{in}^{3})
\end{equation}

The reduction factor $f_{red}$ is represented by

\begin{equation}
    f_{red} = \dfrac{H + 2C_{d} \cdot L }{H + 2C_{d} \cdot L + \frac{3}{2} \cdot \frac{N}{W} \cdot \eta_{max} \cdot A_{rotor}}
\end{equation}

for wind speeds $v_{in}$ above the cut-in velocity $v_{min}$ and below the rated velocity $v_{rated}$ when the turbine output is proportional to the incoming wind speeds (\ref{fig:fig1}b); and 

\begin{equation}
    f_{red} = 1-\dfrac{3}{2} \cdot \dfrac{1}{H + 2C_{d}\cdot L} \cdot \dfrac{H}{L} \cdot \dfrac{ N \cdot P_{el,max}}{J_{in,h}}
\end{equation}

for $v_{in}$ greater than the rated velocity $v_{rated}$, but below the cut-out velocity, $v_{max}$.  For this case, $f_{red}$ is computed only to simulate the effect of wind speed reduction for comparison.  Note that the case of $f_{red} = 1$, KEBA represents the standard approach. 

In these equations for $f_{red}$, $H$ is the height of the boundary layer (Table \ref{table:KEBA}), $C_d$ is the aerodynamic drag coefficient of the surface (Table \ref{table:KEBA}), $L$ and $W$ the length and width of the deployment (Table \ref{table_02}), and $J_{in,h}$ the horizontal kinetic energy flux in the boundary layer ($\rho/2 \cdot v_{in}^3 \cdot W H$).  The values for daytime and nighttime mean boundary layer heights are provided in Table \ref{table:KEBA}. They were derived by comparison of the vertical velocity profiles of the WRF simulations with and without the wind turbine deployment, yielding mean values of about 2000m (day) and 900m (night) (see Supplementary Material).

The kinetic energy budgets for the different scenarios are diagnosed from the time series of the velocity $v_{in}$ and $f_{red}$ and then averaged, with the different terms estimated as in \citeauthor{Kleidon_2020} \citep{Kleidon_2020}.  The budget is defined for the boundary layer air volume enclosing the deployment of wind turbines, given by the dimensions $W$ and $L$ (Table \ref{table_02}), as well as the height of the boundary layer $H$ (Table \ref{table_03}). The magnitude of the budget is set by the influx of kinetic energy, which is determined by the horizontal ($J_{in,h} = WH \cdot \dfrac{\rho}{2} v_{in}^{3}$) and vertical ($J_{in,v} = WL \cdot \rho C_{d} v_{in}^{3}$) influxes of kinetic energy into the volume.  This energy is then either dissipated by surface friction, used for electricity generation, dissipated by wake turbulence, or exported downwind. 

\begin{table}[ht]
\centering
\caption{\label{table:KEBA}Atmospheric and environmental specifications needed for the KEBA estimate.}
\begin{tabular}{@{}lllll}
\hline
\textbf{Description} &\textbf{Symbol} &\textbf{Value} &\textbf{Units} &\textbf{Comments}\\
\hline
 Boundary layer height-Day & $H_{day}$ & 2000 & m & Mean, fixed \\
 Boundary layer height-Night & $H_{night}$ & 900 & m & Mean, fixed\\
 Drag coefficient & $C_{d}$ & 0.001 & - & Mean, fixed\\
\hline
\end{tabular}
\label{table_03}
\end{table}

\section{Results \& Discussion}

We posited that the KE budget is central to capturing the reduction in wind speeds with increased installed capacity to understand the difference between daytime and nighttime yields as shown in Fig. \ref{fig:fig0}, and quantifying the resulting  technical potential.  Therefore, we start by showing the KE budgets during day and night for the different scenarios, the regional reduction in wind speeds, before we describe the estimated yields and capacity factors.  We then perform a sensitivity analysis to boundary layer height to evaluate the effects of the day-night differences and compare these to the general effect of reduced wind speeds with greater installed capacities.  We end this section with a discussion on the limitations, the resulting resource potential estimate, and the broader implicatons.

\subsection{Kinetic Energy budgets}
The KE budget of the boundary layer volume enclosing the deployment is central to KEBA estimates, with the magnitude of the budget defining the wind speed reductions and limiting deployment yields. The horizontal influx accounts for a larger share of the KE budget than the vertical input: 76\% during daytime and 60\% during nighttime. The combination of lower daytime wind speeds ($v_{day,mean}$ = 6.8 $m \: s^{-1}$) and higher boundary heights (H$_{day}$ = 2000m) and higher nighttime wind speeds ($v_{night,mean}$ = 9.5 m s$^{-1}$) and lower boundary layer heights (H$_{night}$ = 900m) lead to similar influxes of kinetic energy of about 150 GW in the mean. The 150 GW budget sets the overall magnitude of the bars in Figure \ref{fig:fig2}(a), with the distribution among the different terms changing due to the different deployment scenarios.

Within the boundary layer volume, KEBA determines the partitioning between the KE influx into frictional dissipation (red), wind turbine yields (dark blue) and wake losses (light blue), and the downwind export of KE out of the deployment volume (light red). KE extracted by wind turbines powers electricity generation ($P_{el,tot}$), with the wakes being dissipated by the mixing behind the turbines ($D_{wake}$). KE extraction consumes KE that would have otherwise been dissipated at the surface by friction or exported downwind. Thus, the increase in capacity density increases yields and wake losses at the expense of KE in downwind export and surface friction. Since individual turbine yields depend on wind speeds, higher nighttime mean wind speeds lead to higher per turbine yield compared to the daytime. Consequently, about ~2\% more KE is extracted by the turbines from the budget at night than during the day (Figure \ref{fig:fig2}(a)).

\subsection{Wind speeds}
\begin{figure}[ht]
\centering
\includegraphics[width = 1.0\textwidth]{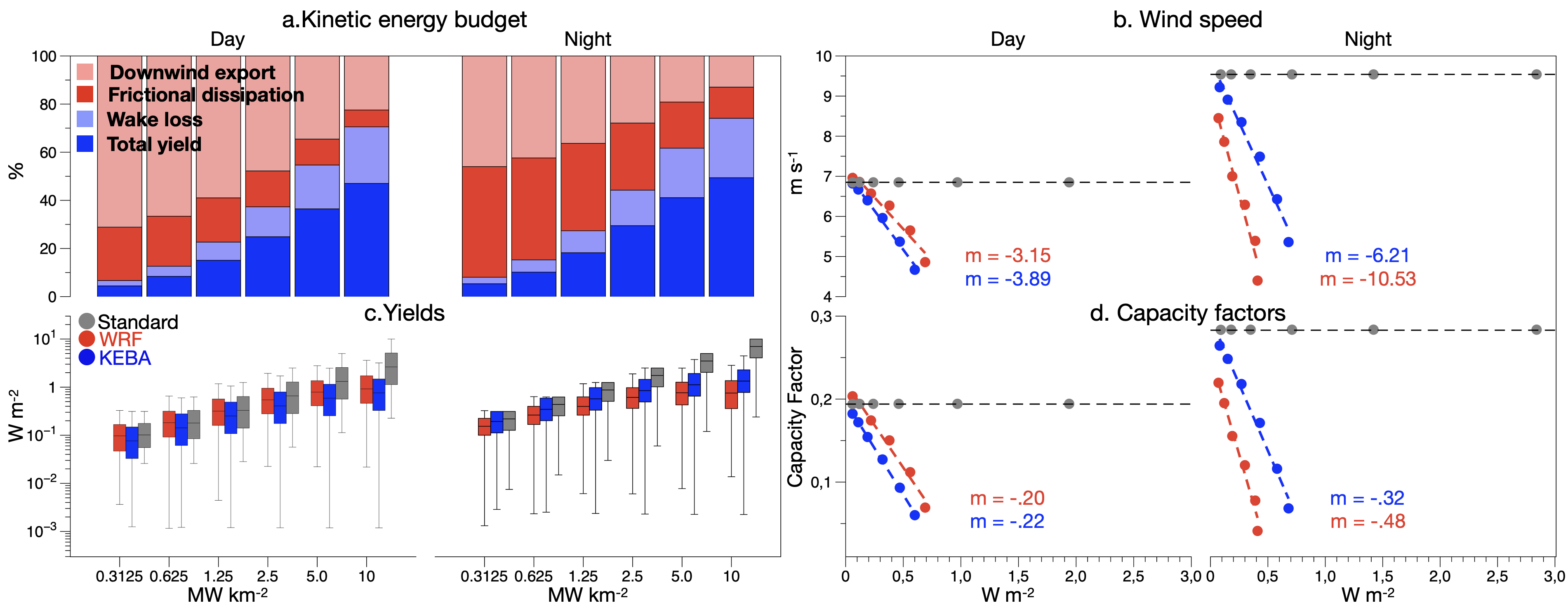}    
\caption{(a) Daytime (left) and nighttime (right) KE budgets, with total yields (dark blue), wake loss (light blue), frictional dissipation (red) and the downwind export (light red). (b) Estimates of wind speeds over the deployment region against the KE extracted by the wind turbines for the "standard" (grey), "WRF" (red) and "KEBA" (blue) estimates. (c) Wind turbine yields as a function of installed capacity density using a logarithmic scale for the "standard" (grey), "WRF" (red) and "KEBA" (blue) estimates. (d)  Capacity factors against the rate of KE extraction for the "standard" (grey), "WRF" (red) and "KEBA" (blue) estimates. Dashed lines denote linear fits, with $m$ values providing the slopes obtained from the linear regression.}
\label{fig:fig2}
\end{figure}

The depletion of the KE budget with increased wind turbine deployment is associated with a reduction in wind speeds. This reduction is shown in Figure \ref{fig:fig2}(b), which shows how the mean wind speed over the deployment region ($v_{eff}$) reduces with the amount of KE extracted by the wind turbines (in W m$^{-2}$ of surface area).  We chose to use the yield on the x-axis rather than installed capacity, because in this way, it shows that wind speed reductions are almost linear with the amount of KE extracted. Figure \ref{fig:fig2}(b) shows these wind speed reductions for the WRF estimate (red) and the KEBA estimate (blue), while the standard approach (grey) assumes no change in wind speeds.  The rates of reduction can be quantified by the slope, $m$, of the linear regressions (dashed lines). Nighttime wind speed reductions ($m_{KEBA,night}=-6.21$) are almost twice as strong as during the day ($m_{KEBA,day}=-3.89$). These reduction rates are similar in the WRF estimates ($m_{wrf,night} = -10.53, m_{wrf,day} = -3.15$). Note that despite the faster rate of reduction in nighttime means, the wind speeds are nevertheless higher in magnitude than during the daytime. Compared to the WRF simulations, KEBA slightly overestimates daytime and underestimates nighttime wind speeds. Thus, the difference in daytime and nighttime wind speed reductions can be directly linked to the lower boundary layer height used in the nighttime KE budget in KEBA.

\subsection{Deployment yields}
Figure \ref{fig:fig2}(c) shows the variation in the wind turbine yields with increasing installed capacity density. Since KEBA models yields as a function of the reduced wind speeds ($v_{eff}$) rather than the prescribed Control wind speeds ($v_{in}$), its estimates (blue) are lower than the standard estimates (gray). KEBA estimates lower additional increments in yields with the increase in installed capacity during both, day and night.  Thus, the diminishing increments in yields with added turbines can be attributed directly to the reduced wind speeds shown in Figure \ref{fig:fig2}(b).  While KEBA estimates of nighttime yields are higher than day, WRF estimates of yield (red) are lower at night than during the day. KEBA captures the trends in yields increments but does not estimate the lower-than-daytime yields at night. It underestimates WRF's mean daytime estimates by 8 - 15\% while overestimating nighttime yields by 20 to 75\%. The standard estimate overestimates yields by up to 180\% during daytime and up to 600\% at night compared to the WRF estimates. The bias in KEBA estimates of yield compared to WRF can be attributed to higher nightime KEBA wind speed estimates. 

\subsection{Capacity Factors}
The lower increments in yields with increased installed capacity indicates that more turbines within the deployment region lowers the mean efficiency of individual turbines.  This can be shown by directly looking at the capacity factors, as displayed in Figure \ref{fig:fig2}(d). Both, KEBA (blue) and WRF (red) estimates show that increasing KE extraction leads to lower capacity factors. The standard estimate (grey), however, assumes no change because no reduction in wind speeds is considered. Again, as in the case for wind speeds, the capacity factors reduce almost linearly with increasing KE extraction.  The slopes of the linear regression show that turbine efficiencies reduce almost twice as fast during the night ($m_{KEBA,night} = -.32$) than during the day ($m_{KEBA,day} = -.22$), which is similar to the WRF estimates ($m_{wrf,night} = -.48$) and ($m_{wrf,day} = -.20$).  While KEBA, again, underestimates the strength of the reduction at night, the close match of KEBA estimates with the WRF estimates highlights that the removal of KE from the boundary layer is the main effect that results in reduced turbine efficiencies and wind turbine yields. KEBA is able to capture a large part of this trend because of the separate definition of day and night KE budgets as opposed to a single KE budget for the whole day.

\subsection{Role of diurnal variations in boundary layer height}
\begin{figure}[ht]
    \centering
    \includegraphics[width=1\textwidth]{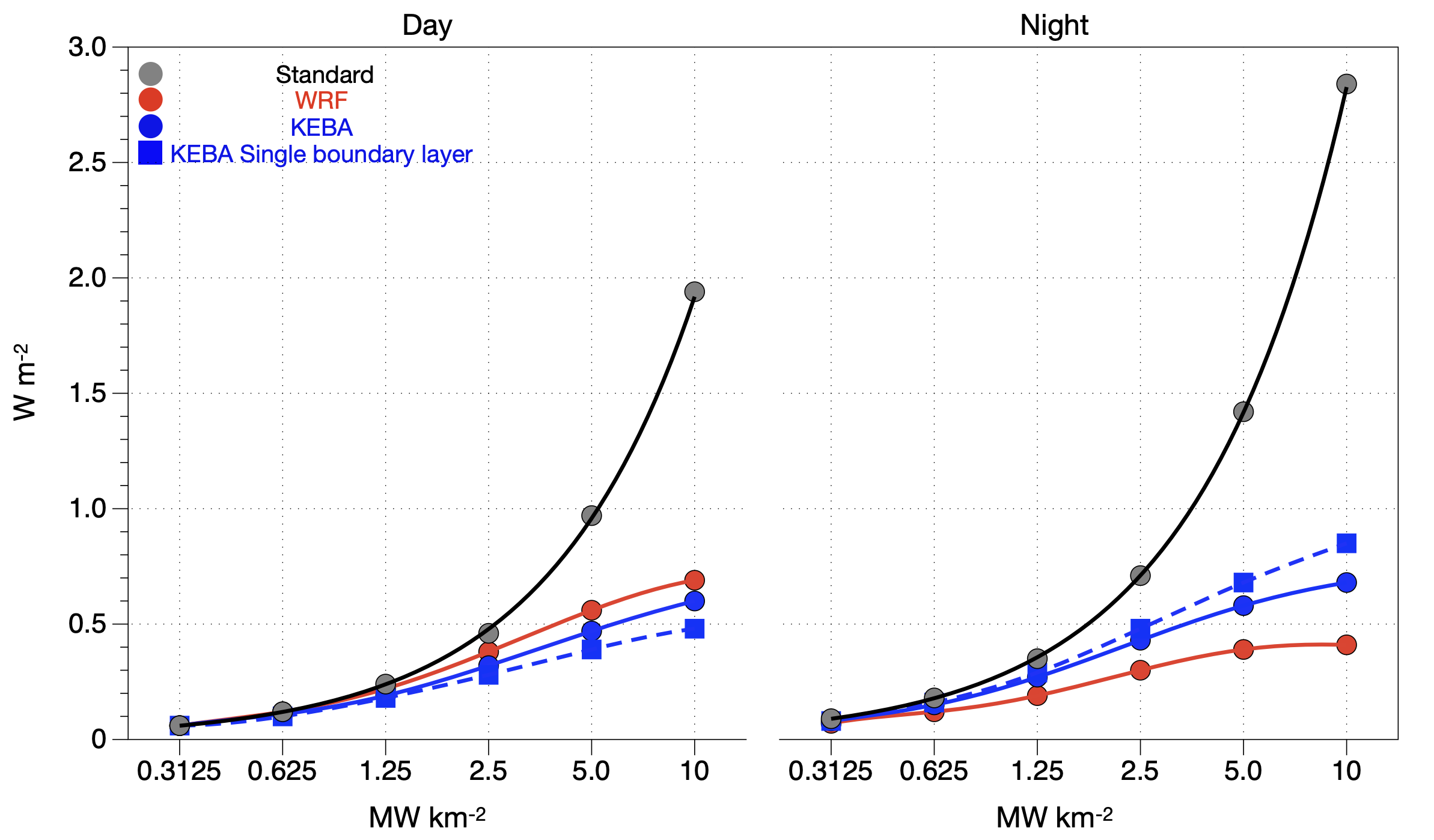}   
    \caption{Daytime (left) and nighttime (right) total yields  estimated by WRF (red $\bigcirc$), KEBA with (blue $\bigcirc$) and without (blue $\Box$ ) diurnal variations in boundary layer height, and the standard approach (grey $\bigcirc$).}
    \label{fig:fig3}
\end{figure}

To evaluate how important the variation in boundary layer height is for estimating yields between day and night, we performed an additional estimate with KEBA in which the boundary layer height is fixed to the mean value of $H = 1268$m (as in \cite{Miller_2015}). This comparison is shown in Figure \ref{fig:fig3}. Although the KEBA estimate with a single mean boundary layer height represents a substantial improvement over the standard estimate, it shows a greater discrepancy to the WRF estimate.  Nighttime yields are overestimated by 20 to 107\% while daytime yields are underestimated by 12 to 31\%. The addition of diurnal variations in boundary layer height improves the estimates relative to WRF estimate, reducing the daytime bias to 10 to 17\% and nighttime bias to 20 to 60\%. The improvement is more pronounced for the nighttime conditions. 

Defining different day and nighttime budgets separately is thus an improvement over neglecting this variation.  It captures more of the underlying mechanism because the daytime solar insolation drives convective motion and higher mean boundary layer heights. The absence of these motions at night lead to much lower boundary layer heights. The difference in the amount of mixing between day and night differentially affects the wind speeds and deployment yields during day and night \citep{Fitch2013b, Akbar_2015}. With all other variables in the KEBA model being fixed, a fixed boundary layer height in KEBA results in a 58\% lower daytime and 30\% higher nighttime KE budget compared to a variable boundary layer height. Although the bias is not entirely compensated for by including the varying boundary layer heights in the KEBA estimates, this information clearly reduces the bias in the direction of the WRF estimate. However, the effect of these diurnal variations at the daily 24 hour scale is relatively muted. This is because the higher day and lower nighttime generations largely compensate for each other implying that it is mainly the role of KE removal that needs to be incorporated in the policy focused estimation of technical potentials. 

\subsection{Limitations}
Although KEBA captures day and night trends produced by WRF better than the standard approach, it is unable to reproduce the lower-than-day, nighttime wind turbine yields from WRF. This is likely because KEBA assumes a well-mixed boundary layer volume that is characterized by one effective wind speed, $v_{eff}$. This assumption is valid during the day when the convective boundary layer is well mixed.  At night, however, stable conditions prevent vertical mixing because the insolation-driven convective motions are absent.  The intensity of mixing within the boundary layer is thus an additional control on the rate at which the KE deficit behind wind turbines is replenished within the boundary layer. The less-mixed nighttime boundary layer slows the replenishment rate, leading to a steeper decline in wind speeds, capacity factors and wind turbine yields \citep{Fitch2013b,Akbar_2015}. 

This interpretation is supported by observations of velocity deficits, or wakes, behind operating offshore wind farms that persist longer when the vertical mixing is lower (55km) than when it is higher (35km)\citep{Cana_2020,Christiansen2005}. Longer wakes during less mixed conditions imply lower downward replenishment than better mixed conditions, leading to slower recovery of wind speeds. 

Reported simulated day- and nighttime mean wind speed reductions of 10 and 30\% \citep{Fitch2013b} from Kansas are similar to the estimates of \citeauthor{Miller_2015} of ~17\% and 43\%. WRF estimates for wind turbine yields during day (~42\% lower than standard) and night (~73\%) are consistent with other simulations of idealized deployment yields over a full diurnal cycle which found that reductions were twice as high at night (57\%) than daytime (28\%)\citep{Akbar_2015}. Thus, it is likely that the differences between WRF and KEBA could be reduced by accounting for stability effects, which could be taken up as a part of future work.

Despite this limitation, KEBA represents a significant improvement over the standard approach, especially with greater installed capacities over the region. Its nighttime yield estimates are within a factor of 2 of the WRF estimate. The standard approach overestimates WRF yields by up to 6 times. Our results highlight the critical role of boundary layer information, in terms of height and mixing/stability, in determining KE budgets that shape the extent to which wind speeds, turbine efficiencies and deployment yields are affected by the removal of KE. Thus, the KEBA estimate appears to be a suitable tool to evaluate Kansas’s technical wind energy potential. \par

\subsection{Reevaluating Kansas's technical potential}

To illustrate the relevance of these KE removal effects, we compare our estimates to the existing technical potentials for Kansas \citep{Lopez_2012,brown_2016}. This comparison is summarised in Table \ref{table:NREL}.  Previous studies estimate potentials of 3101 TWh yr$^{-1}$ and 1877 TWh yr$^{-1}$ for capacity densities of 5 and 3 MW km$^{-2}$ over 1.9 $\cdot$ 10$^{5}$ km$^{2}$ and 1.6 $\cdot$ 10$^{5}$ km$^{2}$, respectively, which include a fixed, 15\% loss in array efficiency. This results in capacity factors of 37\% and 45\%.  Expressed in terms of yields, these estimates imply 1.86 W m$^{-2}$ and 1.36 W m$^{-2}$ of generated electricity per unit surface area.  Multiplied by the deployment areas, these yield technical potentials of 3101 TWh/a and 1877 TWh/a for Kansas in these previous studies.

\begin{table}[ht]
\centering
\caption{\label{table:NREL}Comparison of previously published estimates of the technical wind energy potential of Kansas by \citep{Lopez_2012,brown_2016} with the estimates from this study. For the comparison, we used the scenarios with installed capacity densities of 2.5 and 5 MW km$^{-2}$, which are close to these previous estimates.  A 15\% array efficiency reduction was applied to the Standard estimate.}
\begin{tabular}{@{}lllll}
\hline
 \textbf{Lopez et al. (2012)} &   & Standard & KEBA & WRF \\
 \hline
Deployment area (km$^2$) & 190 000 & & & \\
Capacity density (MW km$^{-2}$) & 5 & 5 & 5 & 5 \\
Capacity factor (\%) & 37  & 41 & 21 & 19 \\
Yield (W m$^{-2}$) & 1.86  & 2.03 & 1.05 & 0.95 \\
Technical potential (TWh yr$^{-1}$)& 3101 & 3379 & 1748 & 1581 \\
Difference (\%) & & +9.0 & -43.6 & -49.0 \\ 
\hline
\textbf{Brown et al. (2016)} &   & Standard & KEBA & WRF \\
\hline
Deployment area (km$^2$) & 157 890 & & & \\
Capacity density (MW km$^{-2}$) & 3 & 2.5 & 2.5 & 2.5 \\
Capacity factor (\%) & 45  & 41 & 31 & 27 \\
Yield (W m$^{-2}$) & 1.36  & 1.02 & 0.75 & 0.68 \\
Technical potential (TWh yr$^{-1}$)& 1877 & 1410 & 1037 & 941 \\
Difference (\%) & & -24.9 & -44.8 & -49.9 \\
\hline
\end{tabular}
\end{table}

We first compare our standard estimate to these resource estimates, using our scenarios with installed capacity densities of 5 and 2.5 MW km$^{-2}$.  For these scenarios, the yield is on average 2.39 W m$^{-2}$ and 1.19 W m$^{-2}$, with a capacity factor of ~48\%. We reduce these estimates by the same 15\% loss, which reduces the yields to 2.03 W m$^{-2}$ and 1.02 W m$^{-2}$ with a 41\% capacity factor.  Multiplied by the deployment areas, these yield technical potentials of 3379 TWh/a and 1410 TWh/a, which are within $\pm 25$\% of the published estimates.

KEBA estimates lower yields of 1.05 W m$^{-2}$ and 0.75 W m$^{-2}$ for the two scenarios, with capacity factors reduced to 21\% and 31\%, respectively.  These reductions compare well with the WRF estimates of 0.95 and 0.68 W m$^{-2}$ and capacity factors of 19\% and 27\% from \citeauthor{Miller_2015}.  Multiplied by the deployment areas, these yield technical potentials of 1748 TWh/a and 1037 TWh/a using KEBA, and 1581 TWh/a and 941 TWh/a using WRF.  These estimates for the technical potentials are lower by 40-50\% due to the reductions in wind speeds. 

Wind speed reductions are thus likely to play a substantial role in lowering regional-scale technical resource potentials than those that use prescribed wind speeds. Note that the potential is nevertheless 3 to 5 times the total energy consumed by the state in 2018 (\cite{kansas_2018_eia}).

\subsection{Implications for technical wind energy potential estimation}

\begin{figure}[ht]
\centering
\includegraphics[width=1\textwidth]{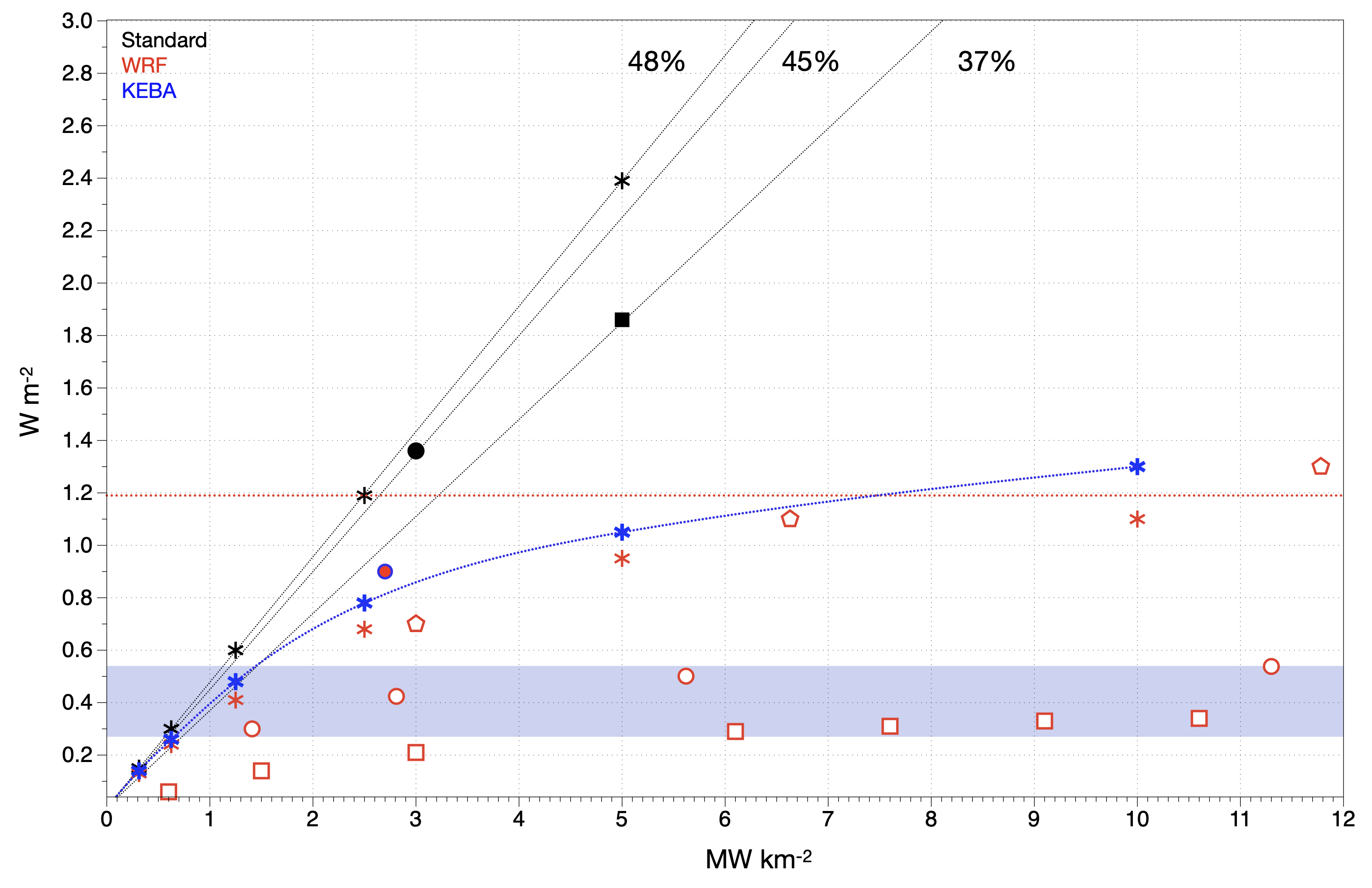}   
\caption{Technical potentials per unit surface area plotted against the capacity density and number of turbines (x-axis). Black symbols represent standard estimates (No KE removal), while red symbols represent meteorological estimates (With KE removal). Blue symbols represent the KEBA estimates from this study. The blue band represents the range of average peak global potentials \citep{Miller_2011,Jacobson2012,Miller_2016, Marvel2012, wang2010potential,wang2010potential,Gustavson1979}. The red dotted line represents the peak average potential of Kansas \citep{Adams_2013} while the dotted lines show the assumed capacity factors without accounting for the removal of KE. Existing estimates of the Kansas resource potential are shown in black filled symbols\citep{brown_2016,Lopez_2012}.The red circle with the blue outline shows an observation-based estimate \citep{Miller_2018}}
\label{fig:fig4}
\end{figure}

The reduced technical potentials derived using KEBA are consistent with previous climate (GCM) and weather modelling (WRF) estimates. This is shown in Fig \ref{fig:fig4} in which the variation of technical potential in Kansas is plotted against the capacity density deployed. KEBA estimates are represented in blue, standard estimates in black and numerical estimates are shown in red. Broadly, the black colour represents the exclusion of KE removal effects while blue and red represent partial (KEBA) and complete inclusion (WRF and GCM), respectively. The stars represent results from this study while the black square \citep{Lopez_2012} and circle \citep{brown_2016} represent previously published estimates of Kansas’s potential. The black dotted lines represent the variation in potential linked to the standard estimates from this study (black stars \citep{Miller_2015}).  The red stars \citep{Miller_2015} and pentagons \citep{Volker_2017} represent previous WRF-based estimates of Kansas. The dotted red line shows the peak average potential from large deployments in Central USA as estimated by \citeauthor{Adams_2013} \citep{Adams_2013}.  Similarly, the red circles \citep{Jacobson2012} and squares \citep{Miller_2016} represent the trends over global land (26\% of global area) derived using GCMs. The blue band highlights the range of peak average global potentials estimated previously \citep{Miller_2011,Jacobson2012,Miller_2016, Marvel2012, wang2010potential,wang2010potential,Gustavson1979}.  The red circle with a blue outline represents an observations - based study of generation from currently operating onshore wind farms in the Central US\citep{Miller_2018}. Numerical estimates in Kansas show that beyond ~1.5 MW km$^{-2}$ of deployed capacity in Kansas leads to ~50\% lower potential (red) compared to the standard estimates. Even though potential increases with increasing capacity deployment, it is not linear as implied by the standard estimates.  This sub-linear increase with installed capacity shows that capacity factors reduce with the increasing deployment. All the numerically simulated estimates display similar variation in potential with capacity and culminate near an average peak of ~1.1 W m$^{-2}$. This variation is also consistent with global estimates over land, albeit higher, because Kansas is winder than most places\citep{Miller_2015}. In line with these estimates, \citeauthor{Miller_2018} \citep{Miller_2018} showed that the actual yield from an average estimated onshore US capacity density of ~2.7 MW km$^{-2}$ is around 0.90 W m$^{-2}$. The agreement between estimates from independent numerical modelling studies and relevant observational data analysis points to the robustness of this trend. The variation in KEBA estimates (blue) is consistent with these trends. Although the match between KEBA and the numerical estimates is not exact, it highlights the significance of the effects of KE removal effects on technical potential.

The comparison in Fig \ref{fig:fig4} shows that the removal of KE is the predominant physical influence which shapes the technical potential of Kansas. The close agreement between KEBA, which only accounts for KE removal effects, and the WRF trends, which includes effects arising from both KE removal and stability, highlights the the role of the KE removal as the predominant influence on technical potentials. The remaining difference indicates the secondary role of stability or the degree of mixing in the boundary layer. 

Including just the KE removal effect leads to a significant improvement in estimates over the standard estimates. Although diurnal variations in stability lead to variations in wind speeds reductions, yields and capacity factors during day and night, the effect of these on the estimated technical potential is marginal over the whole time period. Including additional boundary layer information into KEBA improves the agreement during day and night time but is unable to completely capture stability effects. 

Nevertheless, KEBA still provides a straightforward physical framework through which the role of different physical influences on the resource potential, in this case diurnal variations, can be better understood and quantified. 

\begin{figure}[ht]
\centering
\includegraphics[width=1\textwidth]{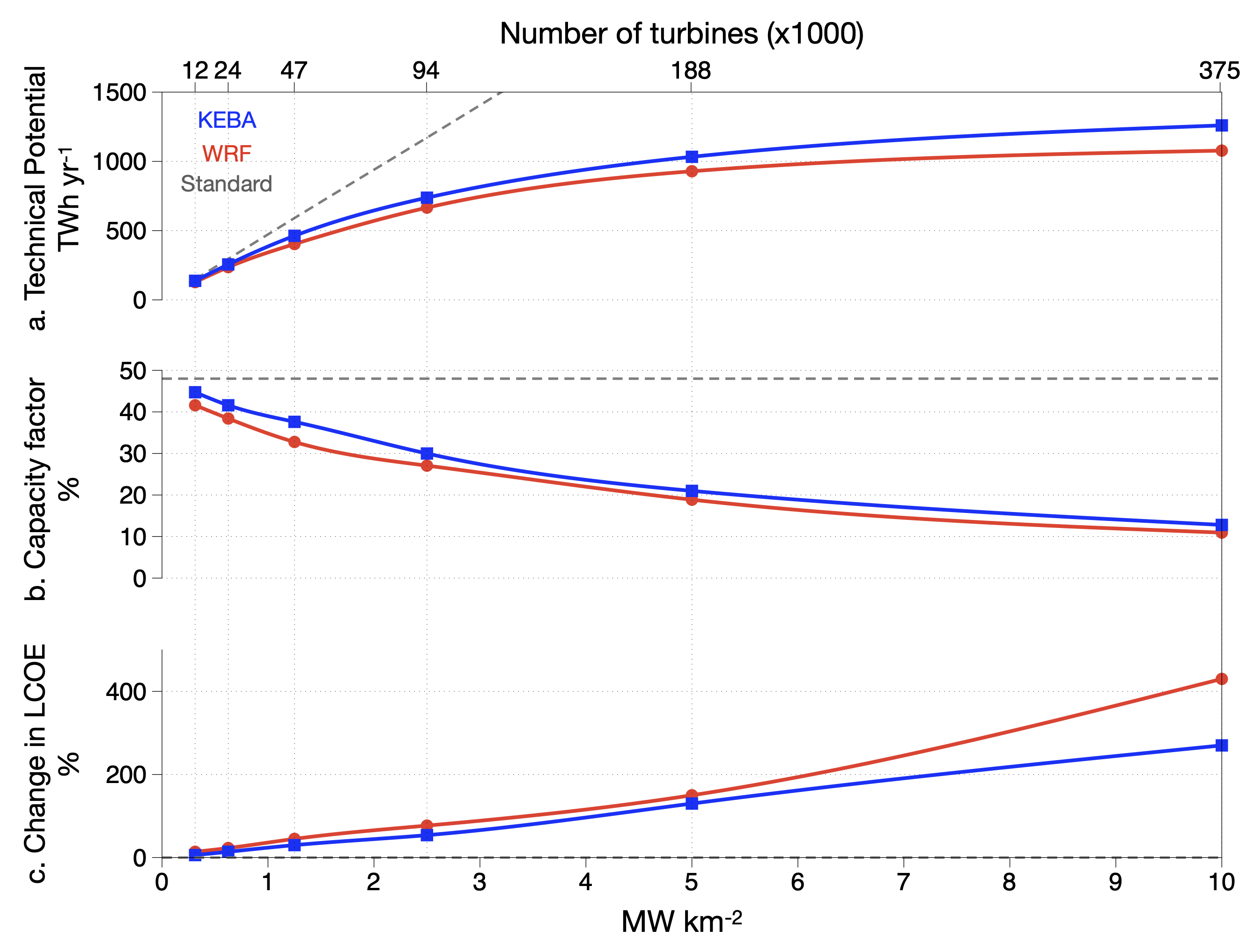}
\caption{(a.) Variation in WRF(red $\bigcirc$) KEBA (blue $\square$) and standard (gray stippled lines) estimates of technical potential,(b.) capacity factors, and (c.) \% change in the Levelised Cost of Energy (LCOE) relative to the standard LCOE estimate plotted as a function of capacity densities (bottom) and number of turbines deployed (top).}
\label{fig:fig6}
\end{figure}

The reduced technical potentials and capacity factors significantly affect the economic potential of wind energy. This is commonly considered by evaluating the economic cost of wind energy using the Levelised Cost of Energy (LCOE) \citep{RAGHEB2017537,Blanco2009}.  We use the estimates from above and plot these in terms of a relative increase in the LCOE in Fig. \ref{fig:fig6}.  

In the standard approach, based on its assumption of constant wind speeds, standard capacity factors remain constant while technical potential increases linearly. A doubling of capacity leads to a doubling of the potential. To estimate LCOE based on standard estimates, we assume that cost of wind energy is only a function of the number of turbines. Then, the LCOE becomes an inverse function of the capacity  (see supplementary materials for details). Thus, there is no change in the standard LCOE as capacity factors remain unchanged (gray stippled line).

In the case of WRF (red circles) and KEBA (blue squares), however, technical potentials increase sub-linearly (Fig \ref{fig:fig6}a) and the capacity factors reduce (Fig \ref{fig:fig6}b). Each doubling of turbines from the lowest scenario to the 2.5 MW km$^{-2}$ scenario leads to an average of 70 - 75\% stepwise increments in potential coupled with an average of 11 - 14\% stepwise reduction in capacity factors. Each doubling in capacity beyond this leads to average stepwise increment of 27 - 31\% in potential coupled with average reductions in capacity factors of 35 - 40\%. Since we assumed that LCOE is only inversely related to capacity factor, reductions in them lead to increases in LCOE, relative to the standard LCOE estimate. Thus, KEBA and WRF lead to estimates of LCOE that are on average  80 - 120\% higher than the standard estimates at an installed capacity density of 5 MW km$^{-2}$. 

This increase in LCOE due to KE removal remains unaccounted for in policy evaluation  because the effect of KE removal on technical potentials and capacity factors is implicitly neglected in the standard approach\citep{Wiser2016}. It should be noted that although we present a simplified illustration, even a detailed LCOE calculation is likely to show similar trends given that LCOE values are highly sensitive to variations in capacity factors \citep{Blanco2009}.  Thus, the reduced potentials arising from KE removal have a significant impact on LCOE which need to be explicitly evaluated in policy evaluations.


\section{Conclusion}
 We conclude that the KE removal effect is the predominant physical influence that shapes technical wind resource potentials at the regional scale. Although day- and nighttime boundary layer heights and stability conditions affect the technical potential, it is the removal of KE from the wind that, primarily, shapes the reduction in wind speeds and capacity factors. It leads to reduced potentials compared to the standard approach that have a significant impact on the economic potential of wind energy at larger scales. 
 
 These impacts need to be assessed in policy evaluations of wind energy and the energy transition. For this KEBA is a viable alternative to the standard approach because it is simple to implement \citep{Kleidon_2020} and accounts for the effect of the key atmospheric control on technical potentials. This is not to negate the use of more physically comprehensive, numerical methods like WRF and GCMs in policy analyses but to enable energy scenario modellers without a background in meteorology to be able to incorporate the key physics without significantly increasing their models’ computational complexity. The heavy computational requirements associated with  physically accurate descriptions of the atmospheric circulations have been reported to inhibit their widespread incorporation into policy side evaluations \citep{Staffel2016}. 
 
Lastly, despite these detrimental effects at larger deployment scales, KEBA's estimates agree with previous research that has shown that wind energy is an abundant and renewable resource that can be harvested to meet a significant part of the future energy demand through efficient, large scale deployment of wind turbines \citep{Jacobson2012, Volker_2017}. 






\bibliography{article}

\newpage
\section*{Supplementary Information}

\setcounter{figure}{0}
\renewcommand{\figurename}{Figure}
\renewcommand{\thefigure}{S\arabic{figure}}

\setcounter{table}{0}
\renewcommand{\thetable}{S\arabic{table}}

\subsection{Determining boundary layer heights for initialising KEBA}
The KEBA model estimates park yield and mean wind speed reduction through the application of conservation of energy (\cite{Kleidon_2020}). The kinetic energy (KE) generated in the atmospheric boundary layer is balanced by that consumed by the wind turbines within the wind park and its wake, dissipated at the surface, and that which powers the remnant wind. The KE conservation is applied to a hypothetical boundary layer volume which encompasses the wind turbine deployment and is mathematically represented as $J_{in,v} + J_{in,h} = P_{el,keba} + P_{wake} + D_{surface} + J_{out,h}$. The left hand side of this equation describes the horizontal and vertical flux of KE in to the boundary layer volume while the right hand side describes how this is partitioned within the volume. The vertical and horizontal KE fluxes into the volume can be expanded in to $J_{in,v} = WL \cdot \rho C_{d} \cdot v_{in}^3$  and $J_{in,h} = WH.\dfrac{\rho}{2} \cdot v_{in}^3$. \par

\begin{figure}[hbt!]
    \centering
    \includegraphics[width=14cm]{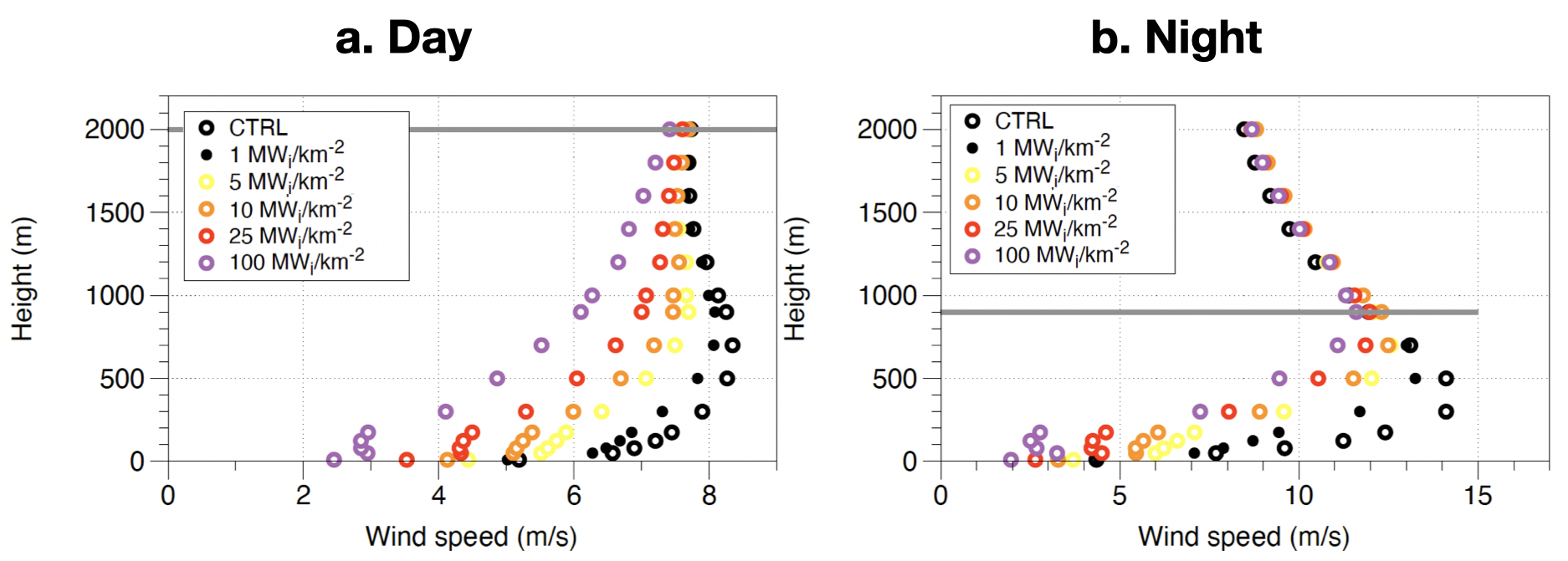}
    
    \caption{Day and night time wind vertical wind speed profiles estimated by \cite{Miller_2015} which show that mean day-time boundary layer height is 2000m whereas that at night is 900m}
    \label{fig:figS1}
\end{figure}

These expressions show that the KE budget available to the wind turbine deployment, is dependent on its geometry (cross-wind width $W$ and downwind length $L$) and the height of the atmospheric boundary layer ($H$). In our analysis, the geometry of the deployment is fixed, therefore the only control on the KE budget is the boundary layer height. Changes in boundary layer height affect the horizontal input of KE flux ($J_{in,h}$). In line with the general definition of the atmospheric boundary layer as the layer which responds quickly to changes in surface forcing (\cite{stull_2009}), the boundary layer from a KE perspective can be also defined as a layer, the kinetic energy content of which responds to changes in surface forcing i.e presence of large wind turbine deployments. Then the boundary  layer height can be understood as the maximum height, up-till which the effects of kinetic energy removal by the turbines can be observed. Since the extraction of KE reduces mean wind speeds, changes in mean wind speeds induced by the turbines can be used estimate this height.

The mean wind speeds over the region of interest, Kansas in this case, were extracted by \cite{Miller_2015} from their WRF simulations. Mean wind speeds were estimated per vertical model level over Kansas for the time period of simulation. Such mean wind speed estimates were computed for all  WRF simulations i.e those without wind parks (CTRL) and those with ($0.3125 - 100 MW_{i} \cdot km^{-2}$). These mean wind speeds from the different models when plotted against model height (m) highlight the vertical variation of mean wind speeds or the vertical wind speed profiles. These are plotted for day and night separately in Figure \ref{fig:figS1}.
In both the plots, the vertical wind speed profiles for the CTRL simulation represent the background circulation in the absence of any wind turbines and hence represent the undisturbed circulation. Vertical wind speed profiles for other simulations deviate from the CTRL trend because  turbines extract KE from the wind speeds, thus slowing them down. The larger the number of turbines within the wind park, the greater is the deviation from the CTRL or the undisturbed trend. The mean day and night boundary layer heights for initialising the KEBA model are then those at which the vertical profiles derived from simulations with wind parks realign themselves with the undisturbed trend. Using this approach Miller et al 2015, estimated the day-time boundary layer height to be 2000m and the night-time boundary layer height to be 900m.

\newpage
\subsection{Wind speed reductions}
Here we tabulate (Table S1) the mean wind speed data simulated by \cite{Miller_2015} without any wind parks or control (CTRL), and with the impact of wind parks with a range of turbine densities ($0.3125 - 10 MW_{i} \: km^{-2}$) split by day and night. Along with it we also provide the mean wind speed reductions estimated by KEBA with different day and night mean boundary layer heights. 
\begin{table}[ht] 
  \begin{center}
    \label{tab:table0}
    \begin{adjustbox}{center}
    \begin{tabular}{c|c|c|c|c|c|c}
     \thead{\textbf{Capacity} \\ \textbf{Density} }&  
     \thead{\textbf{Standard} \\ \textbf{Day} } & 
     \thead{\textbf{WRF} \\ \textbf{Day}} &
     \thead{\textbf{KEBA} \\ \textbf{Day}} &  
     \thead{\textbf{Standard} \\ \textbf{Night}} & 
     \thead{\textbf{WRF} \\ \textbf{Night}} &
     \thead{\textbf{KEBA} \\ \textbf{Night}} \\
     \hline
      $MW_{i} \: km^{-2}$ & $m \: s^{-1}$ & $m \: s^{-1}$ & $m \: s^{-1}$  & $m \: s^{-1}$  &  $m \: s^{-1}$ & $m \: s^{-1}$  \\
      \hline
      0.3125 & 6.85 & 6.96 & 6.85 & 9.54 & 8.45 & 9.22 \\
      0.625 & 6.85& 6.86 & 6.67 & 9.54 & 7.86 & 8.91 \\
      1.25 & 6.85 & 6.57 & 6.40 & 9.54 & 7.00 & 8.35 \\
      2.5 & 6.85 & 6.27 & 5.96 & 9.54 & 6.29 & 7.49 \\
      5.0 & 6.85 & 5.65 & 5.37 & 9.54 & 5.39 & 6.43 \\
      10.0 & 6.85 & 4.86 & 4.67 & 9.54 & 4.40 & 5.36 \\
    \end{tabular}
    \end{adjustbox}
    \caption{This table contains shows wind speed predictions by WRF and KEBA split by day and night. The column titled "standard" represents the CTRL wind speeds i.e with out the impact of reduced wind speeds. Since the standard approach predicts no change to mean wind speeds despite removal of kinetic energy. Thus day and night winds speeds remain constant and same as the CTRL wind speeds.}
    
  \end{center}
\end{table}

\subsection{Park yield and capacity factors}
This section contains tables containing information about park yields and capacity factors estimated by \cite{Miller_2015} (WRF) and by us using the standard approach and the 2 different implementations of KEBA i.e with a single boundary layer height (KEBA single) and another with 2 different average heights (KEBA variable) for day (2000m) and night(900m) for all the capacity density scenarios simulated considered in (0.3125 - 10 MW km$^{-2}$). Tables S2 and S3 contain the data split between day and night time, respectively, whereas Table S4 contains the undifferentiated data.

\begin{table}[H] 
 \begin{center}
    \label{tab:table1}
    \begin{adjustbox}{center}
    \begin{tabular}{c|c|c|c|c|c|c|c|c|c}
     \thead{\textbf{Capacity} \\ \textbf{Density} }& \thead{\textbf{Number} \\ \textbf{of} \\ \textbf{Turbines}} & \thead{\textbf{WRF}} &      \thead{\textbf{Standard}} & 
     \thead{\textbf{KEBA} \\ \textbf{(fixed)}} &
     \thead{\textbf{KEBA} \\ \textbf{(variable)}} &  \thead{\textbf{WRF}\\ \textbf{Capacity} \\ \textbf{Factor}} & 
     \thead{\textbf{Standard}\\ \textbf{Capacity} \\ \textbf{Factor}} &
     \thead{\textbf{KEBA} \\ \textbf{(fixed)}\\ \textbf{Capacity} \\ \textbf{Factor}} &
     \thead{\textbf{KEBA} \\ \textbf{(variable)}\\ \textbf{Capacity} \\ \textbf{Factor}} \\
     \hline
      $MW_{i} \: km^{-2}$ & - & $W_{e} \: m^{-2}$ & $W_{e} \: m^{-2}$  & $W_{e} \: m^{-2}$ &  $W_{e} \: m^{-2}$ & - & - & - & - \\
      \hline
      0.3125 & 11700 & 0.06 & 0.06 & 0.06 & 0.06 & 0.20 & 0.19 & 0.19 & 0.18\\
      0.625 & 23400& 0.12 & 0.12 & 0.10 & 0.11 & 0.19 & 0.19 & 0.16 & 0.17\\
      1.25 & 46800 & 0.22 & 0.24 & 0.18 & 0.19 & 0.17 & 0.19 & 0.14 & 0.15\\
      2.5 & 93600 & 0.38 & 0.49 & 0.28 & 0.32 & 0.15 & 0.19 & 0.11 & 0.13\\
      5.0 & 187200 & 0.56 & 0.97 & 0.39 & 0.47 & 0.11 & 0.19 & 0.08 & 0.09\\
      10.0 & 374400 & 0.69 & 1.94 & 0.48 & 0.60 & 0.07 & 0.19 & 0.05 & 0.06\\
    \end{tabular}
    \end{adjustbox}
    \caption{\textbf{day-time:} This table shows all the capacity density scenarios modelled, associated number of turbines, park yields (WRF) modelled by \cite{Miller_2015}. It also shows the park yields estimated in this study using the standard approach, KEBA with a single boundary layer height (KEBA single) and KEBA with different average bay and night-time boundary layer heights (KEBA variable). The computed capacity factors, represented as fractions, from all the approaches are also included. }
    
  \end{center}
\end{table}

\begin{table}[H] 
  \begin{center}
    \label{tab:table2}
    \begin{adjustbox}{center}
    \begin{tabular}{c|c|c|c|c|c|c|c|c|c}
     \thead{\textbf{Capacity} \\ \textbf{Density} }& \thead{\textbf{Number} \\ \textbf{of} \\ \textbf{Turbines}} & \thead{\textbf{WRF}} & 
     \thead{\textbf{Standard}} & 
     \thead{\textbf{KEBA} \\ \textbf{(fixed)}} &
     \thead{\textbf{KEBA} \\ \textbf{(variable)}} &  \thead{\textbf{WRF}\\ \textbf{Capacity} \\ \textbf{Factor}} & 
     \thead{\textbf{Standard}\\ \textbf{Capacity} \\ \textbf{Factor}} &
     \thead{\textbf{KEBA} \\ \textbf{(fixed)}\\ \textbf{Capacity} \\ \textbf{Factor}} &
     \thead{\textbf{KEBA} \\ \textbf{(variable)}\\ \textbf{Capacity} \\ \textbf{Factor}} \\
     \hline
      $MW_{i} \: km^{-2}$ & - & $W_{e} \: m^{-2}$ & $W_{e} \: m^{-2}$  & $W_{e} \: m^{-2}$ &  $W_{e} \: m^{-2}$ & - & - & - & - \\
      \hline
      0.3125 & 11700 & 0.07 & 0.09 & 0.08 & 0.08 & 0.22 & 0.28 & 0.26 & 0.26\\
      0.625 & 23400& 0.12 & 0.18 & 0.16 & 0.16 & 0.20 & 0.28 & 0.26 & 0.25\\
      1.25 & 46800 & 0.19 & 0.35 & 0.29 & 0.27 & 0.16 & 0.28 & 0.22 & 0.22\\
      2.5 & 93600 & 0.30 & 0.71 & 0.48 & 0.43 & 0.12 & 0.28 & 0.20 & 0.17\\
     5.0 & 187200 & 0.39 & 1.42 & 0.68 & 0.58 & 0.08 & 0.28 & 0.14 & 0.12\\
    \end{tabular}
    \end{adjustbox}
    \caption{ \textbf{night-time:} This table shows all the capacity density scenarios modelled, associated number of turbines, park yields (WRF) modelled by \cite{Miller_2015}. It also shows the park yields estimated in this study using the standard approach, KEBA with a single boundary layer height (KEBA single) and KEBA with different average bay and night-time boundary layer heights (KEBA variable). The computed capacity factors, represented as fractions, from all the approaches are also included. }
    
  \end{center}
\end{table}

\begin{table}[H]
 \begin{center}
    \label{tab:table3}
    \begin{adjustbox}{center}
    \begin{tabular}{c|c|c|c|c|c|c|c|c|c}
     \thead{\textbf{Capacity} \\ \textbf{Density} }& \thead{\textbf{Number} \\ \textbf{of} \\ \textbf{Turbines}} & \thead{\textbf{WRF}} & 
     \thead{\textbf{Standard}} & 
    \thead{\textbf{KEBA} \\ \textbf{(fixed)}} &
     \thead{\textbf{KEBA} \\ \textbf{(variable)}} &  \thead{\textbf{WRF} \\ \textbf{Capacity} \\ \textbf{Factor}} & 
     \thead{\textbf{Standard}\\ \textbf{Capacity} \\ \textbf{Factor}} &
     \thead{\textbf{KEBA} \\ \textbf{(fixed)}\\ \textbf{Capacity} \\ \textbf{Factor}} &
     \thead{\textbf{KEBA} \\ \textbf{(variable)}\\ \textbf{Capacity} \\ \textbf{Factor}} \\
     \hline
      $MW_{i} \: km^{-2}$ & - & $W_{e} \: m^{-2}$ & $W_{e} \: m^{-2}$  & $W_{e} \: m^{-2}$ &  $W_{e} \: m^{-2}$ & - & - & - & - \\
      \hline
      0.3125 & 11700 & 0.13 & 0.15 & 0.14 & 0.14 & 0.42 & 0.48 & 0.45 & 0.45\\
      0.625 & 23400& 0.24 & 0.30 & 0.26 & 0.26 & 0.39 & 0.48 & 0.42 & 0.42\\
      1.25 & 46800 & 0.41 & 0.60 & 0.46 & 0.47 & 0.33 & 0.48 & 0.37 & 0.37\\
      5.0 & 187200 & 0.95 & 2.39 & 1.05 & 1.05 & 0.19 & 0.48 & 0.21 & 0.21\\
      10.0 & 374400 & 1.10 & 4.78 & 1.30 & 1.28 & 0.11 & 0.48 & 0.13 & 0.13\\
    \end{tabular}
    \end{adjustbox}
    \caption{\textbf{Undifferentiated:} This table shows all the capacity density scenarios modelled, associated number of turbines, park yields (WRF) modelled by \cite{Miller_2015}. It also shows the park yields estimated in this study using the standard approach, KEBA with a single boundary layer height (KEBA single) and KEBA with different average bay and night-time boundary layer heights (KEBA variable). The computed capacity factors from all the approaches is also included.}
    
  \end{center}
\end{table}

\subsection{Comparison with published numerical weather model-based estimates of technical wind energy potential}
In Fig \ref{fig:fig4}, we have compared KEBA estimates of technical potential from our analysis with those performed independently by others over comparable regional and global scales using different numerical modelling approaches. For comparison in Kansas and Central USA we used the studies performed by \citeauthor{Adams_2013, Miller_2015} and \citeauthor{Volker_2017} \citep{Adams_2013, Miller_2015, Volker_2017}. From \citeauthor{Volker_2017}, we only use their estimates for their largest deployment scenario (~10$^5$ $km^2$) in Central US. This was the most pertinent case for our analysis. \par
All three of these studies use a version of WRF to model the wind turbine yields and parametrize the wind turbines as momentum sinks. This means that they account for the fact that turbines extract momentum and kinetic energy from the wind thereby lowering wind speeds.While \citeauthor{Miller_2015} and \citeauthor{Adams_2013}use a variation of the  Fitch scheme \citep{Fitch_2013},  \citeauthor{Volker_2017} use the extended wake parameterization or EWP scheme \citep{EWP2015}. The main difference between the Fitch scheme, its variation and the EWP is the while the latter does not include an explicit term to account for the Turbulent Kinetic Energy (TKE) generated by the turbine, the former 2 do. The different schemes lead to differences in the amount of mixing generated withing the boundary layer due to the turbine action. The Fitch scheme estimates more and the EWP relatively less, even though their estimates of wind speeds largely agree with each other\citep{EWP2015}.
\par
It is important to appreciate these differences because \citeauthor{Archer2020}\citep{Archer2020} highlighted two bugs in implementation of the Fitch scheme in WRF versions prior to v4.2 that affect the \citeauthor{Miller_2015} study \citep{Fischereit_2021}.It was shown that the additional term in the Fitch scheme adds excessive TKE  and a coding bug prevents the TKE from being advected properly. Although preliminary analyses have shown that the two errors actually compensate for each other giving rise to TKE estimates that agree with observations \citep{Archer2020,Larsen2021}, it would be useful to briefly evaluate any potential impacts on our results and conclusions.
\par
First, according to a review by \citeauthor{Fischereit_2021} the conclusions of neither of the 3 studies based  used in this study,\citep{Adams_2013, Miller_2015, EWP2015} are affected by the identified bug. Secondly, were these studies affected by the bug or the impact significant one would have expected a more prominent deviation between the Fitch based studies and the EWP based study. This is because the EWP schemes does not use the explicit TKE addition term with which the bug was related. Instead, it is observed that the different studies  exhibit a similar trend of technical potential with installed capacity that culminates to a peak average of ~1.1 W m$^{-2}$.  
\par
Further, the WRF trends in Kansas and Central US are consistent with previous studies that estimate global potentials. Relevant estimates of global land from \citeauthor{Jacobson2012} and \citeauthor{Miller_2016} are shown in Fig \ref{fig:fig6}. These estimates and trends have been derived using global circulation models (GCMs). These are also unaffected by the errors in the Fitch scheme. These trends show the same variation in potentials as the WRF trends i.e. sub linear increase in potential beyond 1.5 MW km$^-{2}$ and culmination to a peak global average range of 0.2 - 0.6 W m$^{-2}$ \cite{Miller_2011,Miller_2016,Jacobson2012,wang2010potential,wang2011potential,Marvel2012}. The agreement between all the independent trends and regional and global scale highlights that the impact of the errors in the Fitch scheme are unlikely  to affect the insights and conclusions generated from this study.

\subsection{Technical Potential, Capacity Factors and Levelized Cost of Energy (LCOE)}

Fig \ref{fig:fig6}b shows that as the number of turbines deployed over the hypothetical wind farm area increases, the removal of kinetic energy (KE) reduces the capacity factor. This means that with the increasing deployed capacity , each turbine produces less energy than what it would have, had it been operating in isolation. The reduction in per turbine efficiency increases with increasing turbines. When the KE removal is neglected, the capacity factor remains unchanged (dotted gray line). While the addition of turbines generally increases the technical potential , the step-wise increments in generation reduce as the turbine numbers increase\ref{fig:fig6}a.  The lower increments are driven by the reductions in capacity factors \ref{fig:fig6}b. 
The effect of this variation in capacity factors can be used to investigate their economic 
impacts using a standard economic cost metric known as the levelized cost of energy or LCOE \citep{RAGHEB2017537}. LCOE is represented by the following formula \citep{RAGHEB2017537}:

\begin{equation}\label{LCOE_full}
    LCOE_{wind} = \dfrac{\sum_{t=1}^{n} (I_t + O\&M_t - PTC_t - D_t + T_t + R_t )\cdot \dfrac{1}{(1+i)^t}}{CF \cdot \sum_{t = 0}^{t = n-1} P_t }
\end{equation}

In this equation, $I_t$ and $O\&M_t$ refer to the capital and operations cost while $PTC_t$, $D_T$, $T_t$ and $R_t$ represent the credits, levies, taxes and royalties, respectively. The term $\dfrac{1}{(1+i)^t}$ is the present value factor which is used to account for the time value of money with a discount factor, $i$, over the lifespan of a wind farm. $t$  represents a year within the operational period of a wind farm. $CF$ is the capacity factor, which in this calculation would be different for different scenarios for WRF and KEBA but same for the standard approach. Since we are interested in simply illustrating only the economic impact of reductions in capacity factors due to KE removal, we can simplify equation \ref{LCOE_full} such that LCOE is only a function of capacity factor. For this, we ignore tax related terms and assume that all costs and installed capacity terms ($P_t$) are sunk and installed at the once at the beginning of the operational life. The time factor also remains constant for all scenarios. It should be noted that this calculation is meant only to illustrate that capacity factor reduction arising from KE removal result in non-trivial increases in LCOE which highlights their inclusion into the policy design.  In reality, turbine installation will occur over many years and so will the cost investments. A real LCOE calculation would need specific and quality controlled inputs about the timing and values of costs, levies and discount rates. With the simplification, equation \ref{LCOE_full} would take the following form.

\begin{equation}\label{LCOE_simple}
    LCOE_{scenario} = \dfrac{1}{CF_{scenario}} \times constant
\end{equation}

In \ref{LCOE_simple}, the LCOE for each capacity density scenario is inversely related to the capacity factor. As the  cost and installed capacity terms are same for the standard and the WRF and KEBA approaches, the percent change relative to the standard approach for each scenario can be calculated. These values for each of the installed capacity density scenarios are plotted for both WRF and KEBA estimates. For example, for the 2.5 MW km$^{-2}$ case the standard approach assumes a 0.48 capacity factor while KEBA and WRF estimates 0.31 and 0.27. Then, to estimate the percent change relative to the standard approach the following approach is used:

\begin{equation}
   \% \, change \, in \, LCOE_{2.5 } = \dfrac{LCOE_{ 2.5,\, KEBA/WRF} - LCOE_{2.5\,, standard}}{LCOE_{2.5\, , standard}}
\end{equation}

These values are tabulated in the table below .  
\begin{table}[H] 
  \begin{center}
    \label{tab:tableLCOE}
    \begin{adjustbox}{center}
    \begin{tabular}{c|c|c|c|c|c|c|c}
     \thead{\textbf{Capacity} \\ \textbf{Density} }& \thead{\textbf{Number} \\ \textbf{of} \\ \textbf{Turbines}} & 
     \thead{\textbf{WRF} \\ \textbf{Capacity} \\ \textbf{Factor}} & 
     \thead{\textbf{Standard}\\ \textbf{Capacity} \\ \textbf{Factor}} &
     \thead{\textbf{KEBA} \\ \textbf{(variable)}\\ \textbf{Capacity} \\ \textbf{Factor}} &
     \thead{\textbf{WRF} \\ \textbf{LCOE} \\ \textbf{Change}} &
     \thead{\textbf{KEBA} \\ \textbf{(variable)} \\ \textbf{LCOE} \\ \textbf{Change}} \\
     \hline
      $MW_{i} \: km^{-2}$ & - & - & - & - & \% & \%\\
      \hline
      0.3125 & 11700 & 0.42 & 0.48 &  0.45 & 14 & 6\\
      0.625 & 23400&  0.39 & 0.48 &  0.42 & 23 & 14\\
      1.25 & 46800 & 0.33 & 0.48 &  0.37 & 45 & 30\\
      2.5 & 93600 & 0.27 & 0.48 &  0.31 & 77 & 54\\
      5.0 & 187200 & 0.19 & 0.48 &  0.21 & 150 & 130\\
    \end{tabular}
    \end{adjustbox}
    \caption{ Tabulation of capacity factors estimated by KEBA, WRF and the standard approach along with the estimated change in LCOE (\%) due to KE removal relative to the standard approach.  }
    
  \end{center}
\end{table}
The change in LCOE is only calculated for the installed capacity range from 0.3125 to 10 MW km$^{-2}$ because this is the range that is typically assumed in wind energy policy scenarios.
They show that as the capacity factor reduces the economic cost of wind energy goes up because each of the turbines performs less efficiently.

\end{document}